\documentclass{article}

\usepackage{nicefrac}
\usepackage{amsmath,amsthm}
\usepackage{amsfonts}

\newtheorem{remark}{Remark}
\newtheorem{definition}{Definition}
\newtheorem{theorem}{Theorem}
\newtheorem{lemma}{Lemma}

\newcommand{\old}[1]{}

\newcommand{\calG}{\mathcal{G}}

\newcommand{\HC}{{H}}

\usepackage{tcolorbox}

\begin{document}

\title{Building a Sybil-Resilient Digital Community Utilizing Trust-Graph Connectivity\thanks{A preliminary version of this paper was presented at the 14th International Computer Science Symposium in Russia, July 1-5, 2019, Novosibirsk, Russia~\cite{communityGrowthCSR19}. 
This current version contains more discussions and clearer representation of theorems and proofs. It also contains two methods for connectivity measurement, while the preliminary version discussed only one.}}

\author{Ouri Poupko%
\thanks{O. Poupko, Weizmann Institute of Science, ouri.poupko@weizmann.ac.il}%
\and Gal Shahaf%
\thanks{G. Shahaf, Weizmann Institute of Science, gal.shahaf@weizmann.ac.il}%
\and Ehud Shapiro%
\thanks{E. Shapiro, Weizmann Institute of Science, ehud.shapiro@weizmann.ac.il}%
\and Nimrod Talmon
\thanks{N. Talmon, Ben-Gurion University, talmonn@bgu.ac.il}}

\maketitle

\begin{abstract}
Preventing fake or duplicate digital identities (aka \emph{sybils}) from joining a digital community may be crucial to its survival, especially if it utilizes  a consensus protocol among its members or employs democratic governance, where sybils can undermine consensus, tilt decisions, or even take over.
Here, we explore the use of a trust-graph of identities, with edges representing trust among identity owners, to allow a community to grow indefinitely without increasing its sybil penetration. Since identities are admitted to the digital community based on their trust by existing digital community members,  \emph{corrupt} identities, which may trust sybils, also pose a threat to the digital community. Sybils and their corrupt perpetrators are together referred to as \emph{byzantines}, and the overarching aim is to limit their penetration into a digital community. We propose two alternative tools to achieve this goal. One is graph conductance, which works under the assumption that honest people are averse to corrupt ones and tend to distrust them. The second is vertex expansion, which relies on the assumption that there are not too many corrupt identities in the community.
Of particular interest is keeping the fraction of byzantines below one third, as it would allow the use of Byzantine Agreement~\cite{lamport1982byzantine} for consensus as well as for sybil-resilient social choice~\cite{shahaf2018reality}.
This paper considers incrementally growing a trust graph and shows that, under its key assumptions and additional requirements, including keeping the conductance or vertex expansion of the community trust graph sufficiently high, a community may grow safely, indefinitely.
\end{abstract}

\section{Introduction}\label{sec: intro}

The goal of this paper is to identify conditions under which a digital community of predominantly \emph{genuine} (singular and unique) digital identities~\cite{DBLP:journals/corr/abs-1904-09630} may grow without increasing the penetration of \emph{sybil} (fake or duplicate) digital identities.  Our particular context of interest is digital democracy~\cite{shapiro2018global,shapiro2018point}, where a sovereign digital community conducts its affairs via egalitarian decision processes; another motivation is the task of growing a permissioned distributed system.
%
Consider an initial digital community with low sybil penetration that wishes to admit new members without admitting too many sybils. As it is not realistic to expect that no sybils will be admitted, the goal is to keep the fraction of sybils below a certain threshold. In a separate paper~\cite{shahaf2018reality}, we show that a digital democracy can tolerate up to one-third sybil penetration and still function democratically.  Still, the fewer the sybils, the smaller the supermajority needed to defend against them.

We model a digital community via a trust graph with a vertex for each identity and with edges representing trust relations between the owners of the corresponding identities (formal definitions are given in Section~\ref{section: prelim}). The model considers \emph{genuine} and \emph{sybil} identities (cf. ~\cite{DBLP:journals/corr/abs-1904-09630}), and refers to the genuine identities that do not trust sybils as \emph{honest} and those that do as \emph{corrupt}.  
Furthermore, to describe an admission process that facilitates incremental community growth, the model presents sequences of trust graphs that may result from such a process.

The goal is to identify sufficient conditions on such graphs, for example, the type of identities in the graph, their relative fractions, and their trust relations, under which a community may grow while keeping the fraction of sybils in it low. 
To achieve this, we use two similar approaches, which differ in the assumptions made on the power of the adversary: The first approach assumes that honest identities tend to trust honest identities rather than corrupt ones, therefore it is hard for the corrupt ones (the adversary) to create trust edges with honest identities.  In this case graph conductance bounds the ratio of sybils in the graph. The second approach assumes that there are not too many corrupt identities, therefore the adversary power is limited by its own size. In this case vertex expansion bounds the ratio of sybils in the graph.

\subsection{Related work}

This section reviews existing work, particularly work that helps clarifying the differences in our proposed model.
  A large portion of the literature on sybil attacks (see, for example, \cite{douceur2002sybil,newsome2004sybil,levine2006survey} and their citations) is focused on \emph{sybil detection}, where the task is to tell the sybil agents from the honest ones. Of particular interest is the approach initiated by Yu et al.~\cite{yu2006sybilguard}, which relies on structural properties of the underlying social network. Yu et al. show how to separate the honest and sybil regions by leveraging the assumption that there are, relatively, few number of edges between them. This framework was studied further~\cite{yu2011sybil,danezis2009sybilinfer,tran2009sybil,tran2011optimal,wei2012sybildefender,cao2012aiding}. 
As pointed out by Alvisi et al.~\cite{alvisi2013sok}, however, such attempts to recover the entire sybil region may potentially occur only in instances where the honest region is sufficiently connected, which is rarely the case in actual social networks. Consequently, Alvisi et al. suggest a more modest goal of producing a whitelist of honest vertices in the graph with respect to a given agent; that is, a \textit{local} sybil detection scheme, in contrast to the \textit{global} ones proposed before. 
Another important aspect of our model that is not apparent in existing works is the differentiation it makes of the identities into three sets (and not merely two): honest, corrupt, and sybil identities.

A problem of a similar flavor is that of \textit{corruption detection} in networks, posed by Alon et al.~\cite{alon2015corruption} and later refined by Jin et al.~\cite{jin2018being}. This setting, inspired by auditing networks, consists of a graph with each of its vertices being either truthful or corrupt, where the overall goal is to detect the corrupt region. In contrast to the sybil detection problem, the corrupt agents are assumed to be immersed throughout the network, and the setting assumes a very restrictive assumption, namely that each agent may accurately determine the true label of its neighbors and report it to a central authority. The authors show how good connectivity properties of the graph allows an approximate recovery of the truthful and corrupt regions.

Note that social networks have some special structure, for example, having low diameters (a.k.a., the \emph{small world phenomena}~\cite{easley2010networks}) or fragmented to highly-connected clusters with low connectivity between different clusters. Moreover, as observed by some researchers~\cite{alvisi2013sok,danezis2009sybilinfer,yu2010sybillimit}, the attacker's inability to maintain sufficiently many attack edges typically results in certain ``bottlenecks'', which can be utilized to pin-point the sybil regions. 

\subsection{Informal Model}

While the problem addressed is related to sybil detection, and indeed we incorporate some of the insights of the works discussed above, here the main goal is different:  Safe community growth.  This work aims to find conditions under which a community may grow without increasing the fraction of hostile members within it; but without necessarily identifying explicitly who is hostile and who is not.  An additional difference from existing literature is the notions of identity and trust. Specifically, existing works consider identities or agents of only two types,  ``good''  and ``bad'', with various names for the two categories. In this work the notion of identities~\cite{DBLP:journals/corr/abs-1904-09630}, is more refined and, we believe, may be closer to reality. 

In particular, this work considers genuine and sybil identities, with the intention that in a real-world scenario these would be characterized by the nature of their \emph{representation}: genuine identities are singular and unique, else are sybil (duplicate or fake, namely not corresponding to a single real person). It further distinguishes between two types of genuine identities, based on their \emph{behavior}: honest, which do not form trust relations with sybils, and corrupt, which do.  This behavioral distinction is captured formally in the proposed model.
We naturally assume that the owners of corrupt identities are the creators and operators of the sybils and that, in the worst case, all sybils and their corrupt perpetrators may cooperate, hence the model labels them together as \emph{byzantines}, and aims to limit their fraction within the community.

We thus begin with a unified formal model of such identities and their  \emph{trust graph}, consisting of vertices that represent identities and edges that represent trust relations among the owners of such identities. The exact definition of these trust relations are outside the scope of this paper, but in a related work~\cite{DBLP:journals/corr/abs-1904-09630} we consider a spectrum of such trust relations, expressed as mutual sureties among identity owners, and inspect their applicability also to the work presented here. Considering the task of sybil-resilient community growth, the model defines the \textit{community history} that aims to capture the incremental changes a community trust graph undergoes in discrete steps. In order to properly characterize identities, the model first employs the basic distinction between genuine and sybil identities. Then, using the community history, it makes a further delicate distinction within genuine identities between \emph{honest} identities, which never trust sybils, and \emph{corrupt} identities, which may trust sybils and, furthermore, may cooperate with other corrupt or sybil community members to introduce sybils into the community.

\begin{figure}[t]
\centering
\includegraphics[width=8.5cm]{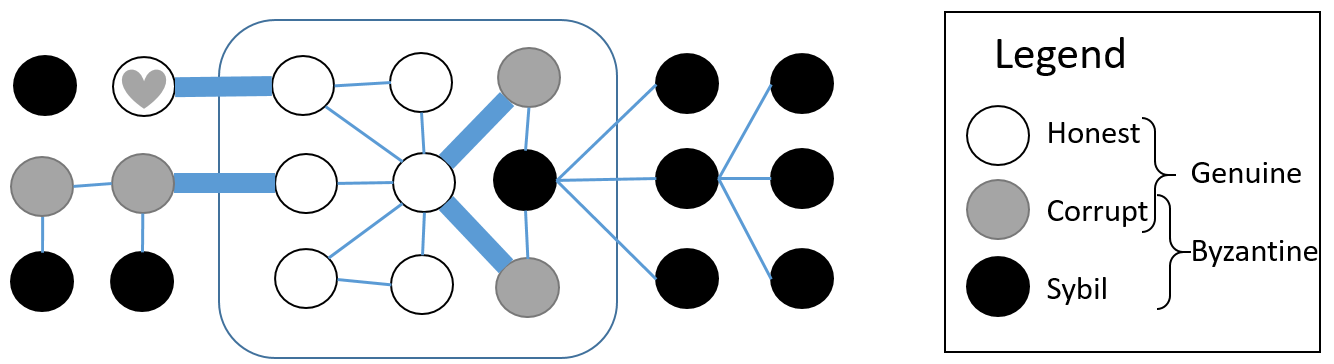}
\caption{Illustration of the general setting:
  The  white vertices (honest identities) \textbf{and} grey vertices (corrupt identities) form the set of genuine identities, while grey vertices (corrupt identities) \textbf{and} black vertices (sybil identities) form the set of byzantines.
  Bold edges represent attack edges. The white vertex with a grey heart in it represents an identity that is ``corrupt at heart'', as currently it does not trust any sybil, but in the future it will (not pictured); thus, the edge connecting it to the honest identity to its right is an attack edge as well. The circled area contains the current community that wishes to grow. Notice that the nine identities in the community contain one sybil and two corrupt identities, thus in particular the community's byzantine penetration is $\beta = \nicefrac{1}{3}$ and the sybil penetration is $\sigma = \nicefrac{1}{9}$.  The fraction  of internal attack edges to the volume of the honest part of the community graph, defined below, is $\gamma_{e}=\nicefrac{1}{8}$.}
\label{figure:one}
\end{figure}

Some assumptions on the power of the sybils and their perpetrators are needed; otherwise there is no hope in achieving our goal.
We present two possible alternative assumptions:
  The first intuitive assumption is that honest identities are averse to corrupt identities, and hence are not likely to trust them. Trust edges that connect honest and corrupt identities are referred to as \emph{attack edges}. So, loosely speaking, the assumption is that there are not too many attack edges. We view this assumption as more realistic than the assumption made in related works~\cite{alon2015corruption,jin2018being}, that truthful agents can identify \emph{precisely} whether a neighbor is corrupt or not. Figure~\ref{figure:one} illustrates the general setting.
The second assumption is that there are not too many corrupt identities in the community. This assumption could be realized, for example, by an incentive mechanism that penalizes for trusting sybils and rewards honest identities. 

\subsection{High-level approach}
\begin{figure}[t]
\centering
\includegraphics[width=8.5cm]{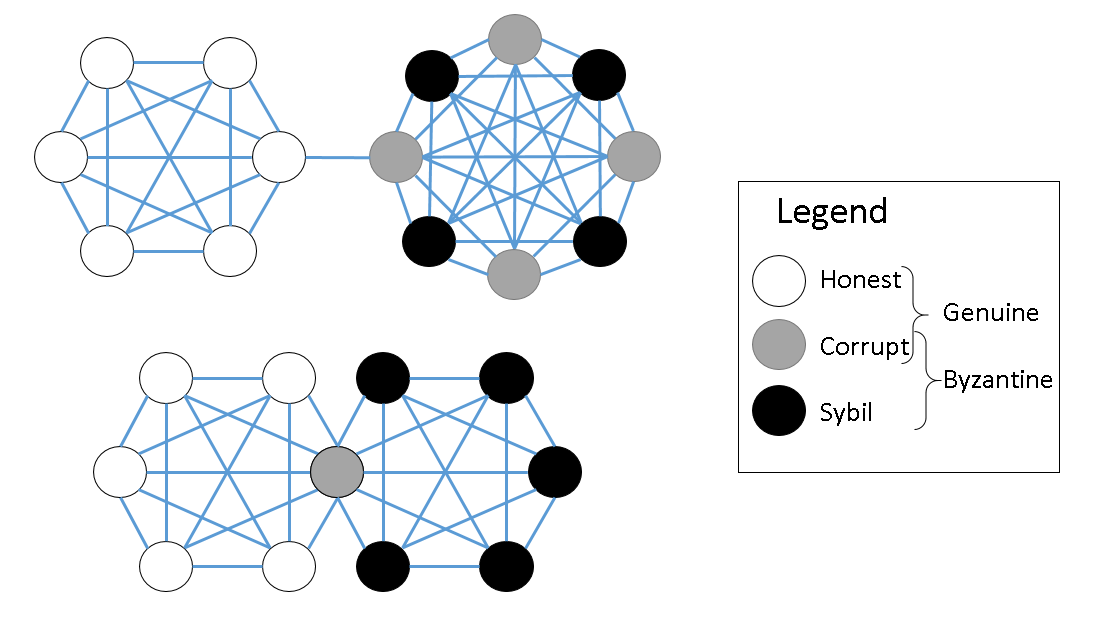}
\caption{Illustration of an extreme example:
  Both community graphs have one cluster of honest identities and one cluster of byzantine identities. In each cluster everyone trusts everyone (the sub graph of the cluster is a clique), yet there is almost no trust between the clusters. The graph at the top demonstrates the case where honest identities don't trust corrupt identities. The graph at the bottom demonstrates the case where there are almost no corrupt identities.}
\label{figure:extreme}
\end{figure}

After defining the three types of population in the community, it is clear that the corrupt identities are the adversary to the goal of growing a community without sybils. Without corrupt identities, if the first identity in the community is not a sybil (therefore it is honest), and given that, by definition, honest identities have no trust edges with sybils, then sybils cannot join the community. To gain intuition regarding the two assumption on the power of adversaries, consider an extreme case, as shown in Figure \ref{figure:extreme}, where the power of the adversary is minimal. The graph on top represents the first assumption, that honest identities are avers to corrupt identities. The graph below represents the second assumption, that there are not too many corrupt identities. In this extreme example  the graph is not constrained in any way, which shows that even a weak adversary can add as many sybils as it wants, without additional measures. Our approach will be to measure the connectivity of the graph and derive a bound on the number of byzantines based on this measurement. The example in Figure \ref{figure:extreme} shows that some simple measurements of connectivity are fruitless for the goal of sybil detection. One such measurement is how dense the graph is, or what is the lower bound on the number of edges within the community. Both graphs show a community where the lower bound on the number of edges is of order $n/2$, and yet the corrupt identities are able to introduce as many sybils as they wish. Another simple measurement is the diameter of the graph, which is also very low in these two communities - 3 at the top and 2 at the bottom.

Yet there is a clear bottleneck in these extreme examples between honest and byzantines. The measures that capture precisely this type of bottleneck are conductance, when the bottleneck is in the edges, and vertex expansion when the bottleneck is in the vertices. The ability to protect the graph from byzantine penetration is based on the key assumption that, while there could be arbitrarily many byzantines wanting to join the growing community, they will have limited connectivity to the current community. Indeed, this observation was applied in the context of sybil detection~\cite{alvisi2013sok,yu2006sybilguard,yu2011sybil,yu2010sybillimit}.

In general, while the connectivity of the whole network is typically fairly low, a social network usually contains many clusters that reflect real life communities. The connectivity of the subgraphs restricted to each of these clusters may be high. In that sense, following Alvisi et al.~\cite{alvisi2013sok}, we adopt a local perspective and focus on the connectivity of the community, regardless of the connectivity of the entire network. In contrast to Alvisi et al.~\cite{alvisi2013sok}, however, we are interested in growing the community and not in whitelisting. Unlike the situation treated by Alvisi et al.~\cite{alvisi2013sok}, which can be viewed as whitelisting, initiated at a singleton community (that is, from a single non-sybil vertex), here we consider arbitrarily-large communities and aim to bound, but not detect or eliminate, the sybils in them.

Specifically, our framework makes use of a ``target conductance'' parameter $\Phi_{e}$, or a ``target vertex expansion'' parameter $\Phi_{v}$, and aims to grow, that is, admit new members, while retaining a conductance of at least $\Phi_{e}$, or $\Phi_{v}$ respectively, at the larger community. Assuming that the initial community harbors a limited attack power and a bounded fraction of byzantines, this paper shows how to safely grow the community, indefinitely. The number of members that may join in each increment is a parameter of the algorithm and is related to the bound on byzantines the community maintains. The lower the bound the more members the community can add in each increment. The bound on byzantines, in turn, depends on the target conductance or vertex expansion that the community maintains. The higher the connectivity of the community, the better the bound on byzantines.

\begin{remark}
Note that our methods are deterministic. That is, they guarantee -- deterministically -- that, if the parameters have certain values and if the assumptions hold, then the conclusion -- namely, that the growing community retains a low fraction of sybil penetration -- holds.
\end{remark}

\subsection{Paper structure}

The paper begins with graph theory terminology and formal definition of graph conductance and vertex expansion in Section~\ref{section: prelim}. For simplicity, the framework describes undirected and unweighted graphs. Note, however, that it may easily be modified and applied to directed and weighted graphs as well. The model is formally described in Section~\ref{sec: model}, by defining types of identities, communities and community history. Then, Section~\ref{section:conductance} describes the first method, based on the assumption of little trust and the use of conductance, and showing sufficient conditions for safe community growth. Section~\ref{section:analysis cond}, shows that the framework is compatible with sparse trust graphs and provides some quantitative estimations of its guarantees. Section~\ref{section:vertex_expansion} and Section~\ref{section:vertex_based_analysis} introduce and analyze the second method, based on the assumption that there are not too many corrupt identities. Section~\ref{section:outlook} concludes with intriguing open questions for future research.

\section{Preliminaries}\label{section: prelim}

This section provides some needed definitions regarding graphs and graph connectivity. Refer to any graph theory textbook, like Diestel's Graph Theory \cite{diestel:graph_theory} for additional background.

Let $G = (V, E)$ be an undirected graph. The \emph{degree} of a vertex $x \in V$ is:
$$\deg(x) := |\{y\in V \> | \> (x,y)\in E\}|$$
$G$ is \emph{$d$-regular} if $\deg(x) = d$ holds for each $x \in V$.
The \emph{volume} of a given subset $A \subseteq V$ is the sum of degrees of its vertices:
$$vol(A) := \sum_{x \in A} \deg(x)$$
Additionally, denote the subgraph induced on the set of vertices $A$ as $G|_A$, the degree of vertex $x \in A$ in $G|_A$ by $\deg_A(x)$, and the volume of a set $B \subseteq A$ in $G|_A$ by:
$$vol_A(B):=\sum_{x\in B} \deg_A(x)$$
Given two subsets $A,B\subseteq V$, the size of the cut between $A$ and $B$ is denoted by:
$$e (A,B) = |\{(x,y)\in E ~|~ x\in A,y\in B\}|$$

\begin{definition}[Conductance]\label{def: conductance}
Let $G = (V, E)$ be a graph. 
The \emph{conductance} of $G$ is defined by:
$$\Phi_{e}(G)= \min_{\emptyset\neq A\subset V} \frac{e(A,A^c)}{\min\{vol(A), vol(A^c)\}}$$
where $A^c := V \setminus A$ is the complement of $A$.
\end{definition}

\begin{remark}
Generally speaking, graph conductance aims to measure the connectivity of the graph by quantifying the minimal cut normalized by the volume of its smaller subset. Conductance should be thought of as the weighted and irregular analogue of edge expansion~\cite{hoory2006expander}, where both notions are essentially equivalent for regular graphs. 
To get a quantitative grip of this measure, notice that for all graphs, $\Phi_{e}\in [0, \frac{1}{2}]$. Intuitively, the conductance of a highly connected graph approaches $\frac{1}{2}$. For example, cliques and complete bipartite graphs satisfy $\Phi_{e}=\frac{1}{2}$, while in a poorly connected graph this measure may be arbitrarily small; for example, a disconnected graph satisfies $\Phi_{e}=0$. 
\end{remark}

The next sections provide theoretical guarantees on sybil safety, given that one can compute conductance. However, determining the exact conductance of a given graph is known to be coNP-hard~\cite{blum1981complexity}. Luckily, the Cheeger inequality \cite{cheeger1969lower} provides a direct relation between conductance of a graph and the second eigenvalue of its random walk matrix, which can be calculated in polynomial time, and approximated in nearly linear time. Refer to ~\cite{hoory2006expander}, ~\cite{jerrum1996markov} and \cite{chung1997spectral} for comprehensive surveys regarding efficient algorithms for measuring conductance. 
 


\begin{definition}[Inner Boundary Vertex Expansion]
Let $G = (V, E)$ be a graph. Given two subsets $A,B\subseteq V$, define the \emph{inner boundary of $A$ w.r.t. $B$} by

$$\partial_{v}(A,B):=\#\{x\in A|\exists y\in B\ s.t.\ (x,y)\in E\}$$

The inner boundary vertex expansion is then defined by:
$$\Phi_{v}(G):=\min_{0<|A|\le\frac{|V|}{2}}{\frac{\partial_{v}(A,A^{c})}{|A|}}$$
\end{definition}

Like conductance, vertex expansion also aims to measure the connectivity of the graph, this time by quantifying the minimal vertex cut, rather than the minimal edge cut. 

To get a quantitative grip of this measure, note that for all graphs $\Phi_{v}\in [0, 1]$. Intuitively, the vertex expansion of a highly connected graph approaches $1$. For example, a clique satisfies $\Phi_{v}=1$, while in a poorly connected graph this measure may be arbitrarily small and a disconnected graph satisfies $\Phi_{v}=0$. Also note the relation between conductance and vertex expansion, given by $\Phi_{v}/d \le \Phi_{e} \le \Phi_{v}$ for $d$-regular graphs.

\section{Formal Model}\label{sec: model}

\subsection{Community Trust Graphs}\label{section:sybils}
The relation between people and their identities is rich and multifaceted. For the purpose of this paper, assume that some identities are \emph{genuine} and others are not, in which case they are called \emph{sybils}. We represent trust relations among identities via a trust graph, in which vertices represent identities and edges represent trust among identities.

\begin{definition}
A \emph{trust graph} $G=(V,E)$ is an undirected graph with vertices that represent identities and edges that represent trust among them.  
\end{definition}

The concept of a community trust graph follows, which depicts the community that grows within such a trust graph.

\begin{definition}
A \emph{community trust graph} $G=(A,V,E)$ is a trust graph with vertices $V$, edges $E$, and a \emph{community} $A \subseteq V$.
\end{definition}

\subsection{Community Histories and Transitions}
The aim of this paper is to find conditions under which a community may grow safely. A graph of identities represents the community. Once establishing some conditions on a given community, we want to verify that these conditions hold under the operation of adding additional identities to the community graph. As the newly added identities threaten these conditions (for example, assume that the community has a bound on the ratio of corrupt identities, and then the added identities may be corrupt and the new community will cross this bound), the model breaks the growth of the community into steps of incremental growth.

\begin{definition}[Community History]
  A \emph{community history} $\calG_V$ over a set of vertices\footnote{As the set of vertices $V$ is fixed in a community history, it does not explicitly model the birth and death of people; modeling this aspect is the subject of future work.} $V$, is a sequence of community trust graphs $\calG_V = G_1,G_2,\ldots$, where $G_i=(A_i,V,E)$, such that $\forall i A_i\subset A_{i+1}$.
\end{definition}

\subsection{Types of Identities}
There are two types of identities: \emph{genuine} and \emph{sybil}. Next, community histories distinguish between two types of genuine identities --  honest and corrupt:  An identity is \emph{corrupt} in a community history if it ever shares an edge with a sybil in this history, and \emph{honest} if it does not. Lumping together sybils and corrupt identities, they form the group of \emph{byzantines}.

The rationale is to bound the number of sybils in the graph, not only at the present but also in the future.  Hence, the model bounds also all potential sybil perpetrators, who may establish trust edges with sybils in the future, in an attempt to introduce them into the community.  Hence, at any point in time (that is, community graph in a community history), a corrupt identity may be only ``corrupt at heart'', with no action as-of-yet to demonstrate its corruption; and the key assumption is that honest identities are averse to corrupt identities even if they are only corrupt at heart.

Below and in the rest of the paper we use disjoint union $A = B \uplus C$ as a shorthand for $A = B \cup C$, $B \cap C = \emptyset$.

\begin{definition}[Types of identities, Attack edges, Sybil penetration]\label{def: identities}
Let~$V$ be a set of vertices that consist of two disjoint subsets $V = T \uplus S$ of genuine $T$ and sybil $S$ vertices, and let $\calG_V$ be a community history over $V$. 
Then, a genuine vertex $t\in T$ is \emph{corrupt} in $\calG_V$  if it trusts a sybil at anytime in $\calG_V$, namely, there is some $(t,s) \in E$, with $t \in T$, $s \in S$, for some $G=(A,V,E) \in \calG_V$. A genuine vertex that is not corrupt is said to be \emph{honest}. Thus, $\calG_V$ partitions the genuine identities $T = H \uplus C$ into honest~$H$ and corrupt $C$ identities.
An edge $(h,c) \in E$ is an \emph{attack edge} if $h\in H$ and $c\in C$.
The \emph{sybil penetration} $\sigma(G)$ of a community trust graph $G=(A,V,E) \in \calG_V$ is $$\sigma(G) = \frac{|A\cap S|}{|A|}$$ 
\end{definition}

\begin{remark}
An important observation is that an attack edge $(h,c)$ may be introduced into a community trust graph in a community history, and be defined as such, even if the corruption of $c$ is still latent in this community trust graph, namely before a trust edge $(c,s)$ between $c$ and a sybil $s$ is introduced.
\end{remark}

In the worst case, sybils and their corrupt perpetrators would cooperate; thus, to allow for incremental community growth, it must bound their combined presence in the community, as defined next:

\begin{definition}[Byzantines and their Penetration]
Let $\calG_V$ be a community history over $V= T \uplus S$ that partitions $T = H \uplus C$ into honest $H$ and corrupt $C$ identities.
Then, a vertex $v \in V$ is \emph{byzantine} if it is a sybil or corrupt and the
\emph{byzantines} $B = S \uplus C$ are the union of the sybil and corrupt vertices.
The \emph{byzantine penetration} $\beta(G)$ of a community trust graph $G=(A,V,E) \in \calG_V$ is $$\beta(G) = \frac{|A\cap B|}{|A|}$$ 
\end{definition}

As $A=(A\cap H)\uplus(A\cap B)$, it would occasionally be convenient to use the equivalence between byzantine penetration to the community $A$ and the fraction of byzantines w.r.t. genuine identities in $A$. Formally,

\begin{equation}\label{eq: relative quotient}
\frac{|A\cap B|}{|A|}\leq \beta \quad \text{iff} \quad 
\frac{|A\cap B|}{|A\cap H|} \leq \frac{\beta}{1-\beta}
\end{equation}

\section{Conductance-Based Approach}\label{section:conductance}

The goal of this section is to find the conditions under which a community can grow while bounding the penetration of byzantines and sybils. The reader may read the following remedy as high level instructions to achieve this goal:
\begin{enumerate}
    \item Start with an initial community.
    \item Choose the desired bound on byzantine penetration.
    \item Measure the fraction of edges within the community, out of all edges stemming out of the community.
    \item Estimate a bound on the connectivity between honest and sybil/byzantine identities.
    \item Admit new candidates to the community only if the connectivity within the target community is sufficiently large.
\end{enumerate}

The following provides sufficient conditions for byzantine-resilient community growth, under the assumption that honest people tend to trust honest people and distrust corrupt people. 

\begin{theorem}\label{thm: safety}
Let $\calG_V$ be a community history. Set parameters $\alpha \in [0,1],\beta\le\frac{1}{2}-\frac{1}{|A_1|},\gamma_e\in[0,\frac{1}{2}],\delta=1-2\beta$. Assume:
\begin{enumerate}
    \item All communities have a bounded degree, both above and below:
    $$\alpha\cdot d\le deg_{A_i}(v)\le d \text{ for all } v\in A_i\ , i\in \mathbb{N}$$
    
    \item Byzantine penetration to the initial community is bounded:
    $$\beta(G_1)\le\beta$$
    
    \item The edges between honest and byzantine identities are relatively scarce: 
    $$\frac{e(A_{i}\cap \HC,A_{i}\cap B)}{vol_{A_{i}}(A_{i}\cap \HC)}\leq \gamma_e$$
    
    \item Community growth is bounded:
    $$|A_{i}\setminus A_{i-1}| \leq \delta |A_{i-1}|$$
    
    \item The conductance within $A_i$ is sufficiently high:
    $$\Phi_{e}(G|_{A_{i}}) > \frac{\gamma_e}{\alpha}\cdot \left(\frac{1-\beta}{\beta}\right)$$
    
\end{enumerate}

Then, every community $G_i\in \calG_V$ has Byzantine penetration $\beta(G_i) \le \beta$.
\end{theorem}

Roughly speaking, Theorem \ref{thm: safety} suggests that whenever: (1) Each graph $G_i|_{A_{i}}$ has a bounded degree, both above and below; (2) Byzantine penetration to $A_{1}$ is bounded; (3) Edges between honest and byzantine identities are scarce; (4) Community growth in each step is bounded; (5) The conductance within $G_i|_{A_{i}}$ is sufficiently high; Then, the community may grow indefinitely with bounded byzantine penetration. 

Theorem \ref{thm: safety} follows by induction from the following Lemma:

\begin{lemma}\label{lemma: cond main}
Let $G=(A,V,E)$ and $G'=(A',V,E)$ be two community trust graphs, where $A\subset A'$. Set parameters $\alpha\in[0,1]$ and $\beta,\gamma,\delta \in [0,\frac{1}{2}]$. Assume:

\begin{enumerate}
    \item Each vertex in $A'$ has a bounded degree, both above and below:
    $$\alpha\cdot d\le deg_{A'}(v)\le d \forall v\in A'$$

    \item Byzantine penetration to the initial community is bounded:
    $$\beta(G)+\frac{\delta}{2}\le\frac{1}{2}$$

    \item The edges between honest and byzantine identities are relatively scarce:
    $$\frac{e(A'\cap \HC,A'\cap B)}{vol_{A'}(A'\cap \HC)}\leq \gamma_e$$

    \item Community growth is bounded:
    $$|A'\setminus A| \leq \delta |A|$$

    \item The conductance within $A'$ is sufficiently high:
    $$\Phi_{e}(G|_{A'}) > \frac{\gamma_e}{\alpha}\cdot \left(\frac{1-\beta}{\beta}\right)$$

\end{enumerate}
$\>\>\>$ Then, $\beta(G')\le\beta$.
\end{lemma}

\begin{proof}
First note that even if all the added identities from $A$ to $A'$ are byzantines, it still follows that

\begin{equation*}
    |A'\cap B| \le |A\cap B|+|A'\setminus A|=\beta(G)\cdot|A|+|A'|-|A|
\end{equation*}

Applying assumption (2):
\begin{equation*}
    |A'\cap B| \le \frac{(1-\delta)|A|}{2}+|A'|-|A|=\frac{|A'|}{2}-\frac{\delta|A|}{2}+\frac{|A'|}{2}-\frac{|A|}{2}
\end{equation*}

Applying assumption (4):
\begin{equation*}
    |A'\cap B|\le\frac{|A'|}{2}-\frac{\delta|A|}{2}+\frac{\delta|A|}{2}=\frac{|A'|}{2}
\end{equation*}

As $V=B\uplus H$, it follows that: 
\begin{equation} \label{eq: B<H cond}
    |A'\cap B|\le|A'\cap \HC|   
\end{equation}

Now utilizing assumption (1):
\begin{align}\label{eq: vol_A'1}
    vol_{A'}(A'\cap B) &:= \sum_{a\in A'\cap B} |\{x\in A'~|~(a,x)\in E\}| \nonumber \\ 
    &\ge \sum_{a\in A'\cap B} \alpha d = \alpha d|A'\cap B|\ .
\end{align}

Similarly, the following holds:

\begin{equation} \label{eq: vol_A'2}
    vol_{A'}(A'\cap \HC)\ge\alpha d|A'\cap \HC|
\end{equation}

Inequalities \ref{eq: B<H cond} and \ref{eq: vol_A'2} imply that:

$$vol_{A'}(A'\cap \HC)\ge\alpha d|A'\cap B|$$

and together with Inequality \ref{eq: vol_A'1}:

\begin{equation}\label{eq:min vol}
    \min\{vol(A'\cap \HC), vol(A'\cap B)\} 
    \ge \alpha d|A'\cap B|
\end{equation}

Now, Inequality \ref{eq:min vol} and assumption (5) imply that:
\begin{align*}
\begin{split}
    \frac{e(A'\cap \HC,A'\cap B)}{\alpha d|A'\cap B|}&\ge\frac{e(A'\cap \HC,A'\cap B)}{\min\{vol(A'\cap \HC), vol(A'\cap B)\}} \\
    &>\frac{\gamma_e}{\alpha}\cdot \left(\frac{1-\beta}{\beta}\right)\ ,
\end{split}
\end{align*}

or equivalently:
\begin{equation}\label{eq: conductance result}
    \frac{e(A'\cap \HC,A'\cap B)}{ d\gamma_e |A'\cap B|} \ge \frac{1-\beta}{\beta}
\end{equation}

Assumptions (1) and (3) imply:

\begin{equation*}
\frac{e(A'\cap \HC,A'\cap B)}{d|A'\cap \HC|}\leq\frac{e(A'\cap \HC,A'\cap B)}{vol_{A'}(A'\cap \HC)}\leq \gamma_e
\end{equation*}

or equivalently:

\begin{equation}\label{eq: gamma result}
    |A'\cap \HC| \ge \frac{e(A'\cap \HC,A'\cap B)}{d\gamma_e}
\end{equation}

Combining Inequalities \ref{eq: conductance result} and \ref{eq: gamma result}:
\begin{align*}
\frac{|A'|}{|A'\cap B|}&=\frac{|A'\cap \HC|+|A'\cap B|}{|A'\cap B|}\\
&\ge\frac{e(A'\cap \HC,A'\cap B)}{d\gamma_e|A'\cap B|}+1\\
&>\left(\frac{1-\beta}{\beta}\right)+1=\frac{1}{\beta}\ ,
\end{align*}
where the first equality holds as $A=(A\cap \HC)\uplus(A\cap B)$, the second inequality stems from Equation \ref{eq: gamma result} and the third inequality stems from Equation \ref{eq: conductance result}. Flipping the nominator and the denominator then gives $\beta(A'):=\frac{|A'\cap B|}{|A'|}<\beta$.
\end{proof}

\begin{remark}
    A potential application of lemma \ref{lemma: cond main} is a byzantine-resilient union of two communities. Let $A,A'\subseteq V$ denote two communities that have some overlap (non-empty intersection) and wish to unite into $A_2 := A \cup A'$.  Then, if lemma \ref{lemma: cond main} holds for $(A_1, A_2)$ in case $A_1 := A$ and also in case $A_1 := A'$, this would provide both $A$ and $A'$ the necessary guarantee that the union would not result in an increase of the sybil penetration rate for either community.
\end{remark}

\section{Analysis of the Conductance-Based Approach}\label{section:analysis cond}

Our results show the conditions under which a community can grow and maintain sybil safety. It is still not clear however if such conditions are practical. This section takes a closer look at graphs, graph conductance and the interplay between the parameters. We show that under the range of possible parameters in the model and the required conductance derived from these parameters there are indeed many such graphs that meet the requirements. Theoretically, a fully connected graph easily holds these requirements, but trust graphs are rather sparse graphs, so specifically the question is whether sparse graphs can hold these requirements.

\subsection{Sparse Graphs}
Recall that the safety of the community growth, more specifically the required level of conductance for the community to grow safely, relies upon the parameters $\alpha$, $\beta$, and $\gamma_{e}$. While a given community may evolve wrt.\ any choice of parameters, some choices will inevitably yield degenerate outcomes; one case is as the model requires $\Phi_{e}(G|_{A'}) > \frac{\gamma_{e}}{\alpha}\cdot \left(\frac{1-\beta}{\beta}\right)$, while the conductance of any graph is upper bounded by $\frac{1}{2}$. Specifically, whenever $\gamma_{e}\left(\frac{1-\beta}{\beta}\right)>\frac{1}{2}$, the community cannot possibly grow, regardless of the choice of $\alpha$.
While complete graphs and complete bipartite graphs are the classic examples of graphs which satisfy $\Phi_{e}(G|_{A'}) = \frac{1}{2}$, the fact that their degree is of order $d=\Theta(n)$ makes them unrealistic in our setting, where agents may potentially trust only a uniformly-bounded number of identities. In this context, the main question seems to be the following: \emph{Could a given community safely grow while retaining a given maximal degree $d$?}
Surprisingly, not only that the answer is affirmative, it also holds for a plethora of trust graphs. We utilize Friedman's classical result:

\begin{theorem} (Friedman \cite{friedman2003proof}, rephrased)
  Let $G$ be a random $d$-regular graph on $n$ vertices. Then, for any $0<\epsilon$,
  $\lambda(G)\leq \frac{2\sqrt{d-1}}{d}+\epsilon$ holds with probability $1-o_n(1)$.
\end{theorem}

Thus, almost all $d$-regular graphs on $n$ vertices satisfy $\lambda_2\leq \frac{2}{\sqrt{d}}$. Applying this term in Cheeger's inequality yields that such graphs satisfy 
\begin{equation}\label{eq: conductance vs. degree}
\frac{1}{2}-\frac{1}{\sqrt{d}}\leq \Phi_{e}
\end{equation}
meaning that the choice of $d$ affects the level of conductance one hopes to achieve.

\subsection{Parameter Interplay}\label{subsection:edge_param_interplay}
The following subsection considers numerical examples to better appreciate the analysis above.
First, consider the realistic assumption where each identity is assumed to trust up to $d=100$ identities (notice that this can be enforced by the system). Equation~\ref{eq: conductance vs. degree} now suggests that a random graph of degree $d$ on $n$ vertices (where $d$ may be constant wrt.\ $n$) satisfies $\Phi_{e} >\frac{2}{5}$. For simplicity, we take this quantity as a benchmark. It follows that whenever $\frac{\gamma_{e}}{\alpha}\cdot \left(\frac{1-\beta}{\beta}\right)<\frac{2}{5}$, there exist a plethora of potential community histories for which a given community may potentially grow to be arbitrarily large.
Some further examples:
\begin{enumerate}
    \item If $\gamma_{e} = 0$, then any community history that begins with a connected byzantine-free community would retain $0$-byzantine penetration;
    \item The choice $\beta = 0$ is not attainable, corresponding  to the intuition that one can never guarantee a completely byzantine-free community growth. \end{enumerate}

\begin{figure}[t]
%
  \centering
        \includegraphics[width=6cm]{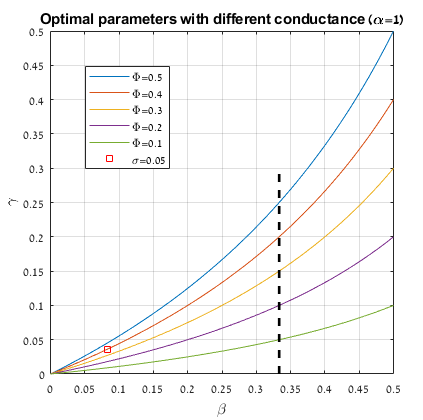}
        \includegraphics[width=6cm]{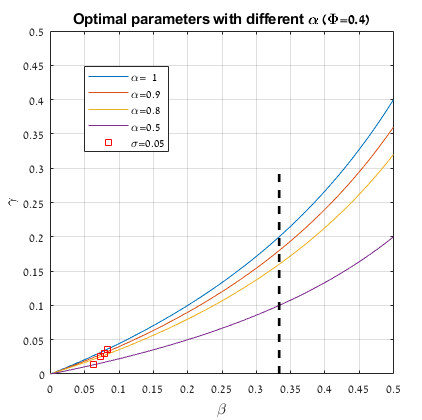}
  \caption{
Parameter Interplay. The plot on the left shows $\gamma_{e}$ as a function of $\beta$, for $\alpha = 1$, where each line represents a different conductance $\Phi_{e}$ value. It shows, for example, that if the community fixes $\alpha = 1$ and sets $\Phi_{e} = 0.4$, then to achieve $\beta = 0.2$ it can tolerate $\gamma_{e} = 0.1$. The plot on the right shows the effect of $\alpha$, for $\Phi_{e} = 0.4$. In both plots, the red rectangles show respective $\beta$ and $\gamma_{e}$ values ensuring $\sigma = 0.05$.}
    \label{figure:interplay}
%
\end{figure}

Figure~\ref{figure:interplay} illustrates the parameter interplay further.
Notice that the key assumption, stating that honest people tend to trust honest people more than they tend to trust corrupt people, implies that $\gamma_{e} < \beta$ (as $\gamma_{e} > \beta$ implies that honest people trust corrupt people more than their relative share in the community).\footnote{In a separate line of research (in preparation) we consider processes and mechanisms that help lowering $\gamma_{e}$ even further.}

\subsection{Parameter Estimation}\label{section:quantitative}
While $\alpha$ and $\Phi_{e}$ can be decided by the community (either by the foremothers of the community or by a global, decentralized democratic decision making process), $\beta(G)$ and $\gamma_{e}$ rely on the dynamics of the community history. To incrementally grow the community at a given time, one may settle for estimating the current state of affairs, as follows.
Specifically, assuming that a thorough examination of a given identity could determine whether it is genuine or sybil, one may apply random checks to empirically estimate $\beta(G)$ and $\gamma_{e}$. This could be carried out in the following manner:\footnote{A related sampling-based approach to estimate the number of sybils is briefly discussed by Shahaf et al.~\cite[Remark 2]{shahaf2018reality}.}
\begin{enumerate}
    \item Examination of an identity $x\in V$ determines whether it is genuine or sybil
    \item Examination of the neighbors of a genuine identity $x\in V$ (the ball of radius~$1$ around it) determines whether it is explicitly (but not latently) corrupt
    \item Examination of the ball of radius $2$ around an honest identity $x$ determines whether its neighbors are explicitly byzantine
\end{enumerate}


\section{Vertex Expansion Approach} \label{section:vertex_expansion}

The next section presents our second assumption, which focuses on the corrupt identities themselves, rather then the trust between honest identities and corrupt identities. Thus, we simply assume that there is a bound on how many identities in a community are corrupt. In a trust graph this results in a limited number of vertices on the boundary between honest identities and sybil identities. The following provides sufficient conditions for byzantine-resilient community growth, under the assumption that the population of corrupt identities in the community is bounded. This time we use vertex expansion to derive a bound on the number of byzantine identities.

\begin{theorem}\label{thm: safety ve}
Let $\calG_V = G_1, G_2, \ldots$ be a community history over $V$. Let $\beta\le\frac{1}{2}-\frac{1}{2|A_1|}$, $\gamma_{v}\in [0,\frac{1}{2}]$, and $\delta=1-2\beta$. Assume:

\begin{enumerate}
    \item Byzantine penetration to the initial community is bounded:
    $$\beta(G_1)\le\beta$$

    \item The population of corrupt identities is bounded:
    $$\frac{|A_i\cap C|}{|A_i|}\leq \gamma_v$$

    \item Community growth is bounded:
    $$|A_{i}\setminus A_{i-1}| \leq \delta |A_{i-1}|$$

    \item The vertex expansion within $A_i$ is sufficiently high:
    $$\Phi_{v}(G|_{A_{i}}) > \frac{\gamma_v}{\beta}$$

\end{enumerate}

Then, every community $G_i\in \calG_V$ has Byzantine penetration $\beta(G_i) \le \beta$.
\end{theorem}

Notice that there is one less parameter $\alpha$ in the vertex based version of the model. While it was required in the edge based version, to establish a lower bound on the volume of $H$, and although it has a strong intuition for our goal (the more honest identities trust each other, the harder it is for the untrusted to penetrate their community), the theorem for the vertex based version will hold without it. This makes this version slightly simpler, as there is one less parameter that the community needs to decide upon.

As before, theorem \ref{thm: safety ve} follows by induction from the following Lemma:

\begin{lemma}

Let $G=(A,V,E)$ and $G'=(A',V,E)$ be two community trust graphs, where $A\subset A'$. Set parameters $\beta,\gamma,\delta \in [0,\frac{1}{2}]$. Assume:

\begin{enumerate}
    \item Byzantine penetration to the initial community is bounded:
    $$\beta(G)+\frac{\delta}{2}\le\frac{1}{2}$$

    \item The population of corrupt identities is bounded in $A'$:
    $$\frac{|A'\cap C|}{|A'|}\leq \gamma_v$$

    \item Community growth is bounded:
    $$|A'\setminus A| \leq \delta |A|$$

    \item The vertex expansion within $A'$ is sufficiently high:
    $$\Phi_{v}(G|_{A'}) > \frac{\gamma_v}{\beta}$$

\end{enumerate}
$\>\>\>$ Then, $\beta(G')\le\beta$.
\end{lemma} \label{theorem: vertex_main}

\begin{proof}
Similarly to the proof of lemma \ref{lemma: cond main}, assumptions (1) and (3) imply that:

\begin{equation} \label{eq: B<H ve}
    |A'\cap B|\le|A'\cap \HC|
\end{equation}

Inequality \ref{eq: B<H ve} and assumption (4) imply that:

$$\frac{\gamma_{v}}{\beta}\leq\Phi_{v}(G'|_{A'})\leq\frac{\partial_{v}(A'\cap B,A'\cap H)}{|A'\cap B|}\leq\frac{|A'\cap C|}{|A'\cap B|}$$
where the last inequality stems from definition \ref{def: identities} (there are no edges between $H$ and $S$, therefore the boundary between $B$ and $H$ is a subset of $C$). Applying assumption (2) it follows that:

$$\frac{\gamma_{v}}{\beta}\leq\frac{\gamma_{v}|A'|}{|A'\cap B|}$$

which leads to

$$\frac{|A'\cap B|}{|A'|}\le\beta$$

That is, $G'$ has byzantine penetration $\beta(G') \le \beta$.
\end{proof}

\begin{remark}
Our two results for community growth, one based on conductance and the other based on vertex expansion, are very similar. The main difference between them lies in the premises of the two corollaries. The first assumes that honest people tend to trust honest people more than they tend to trust corrupt people. The second, which may be more na\"{i}ve, directly assumes that there are not too many corrupted people in a given community to begin with. Again, the conditions under which we assume either of these bounds to be low is the subject of a separate line of work.
\end{remark}

\section{Analysis of the Vertex Expansion Approach}\label{section:vertex_based_analysis}
\begin{figure}[t]
%

  \centering
  \includegraphics[width=6cm]{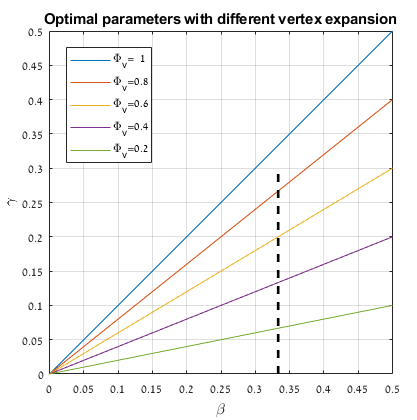}
  \caption{
Parameter Interplay. The plot shows $\gamma_{v}$ as a function of $\beta$, where each line represents a different vertex expansion $\Phi_{v}$ value. It shows, for example, that if the community sets $\Phi_{v} = 0.6$, then to achieve $\beta = \frac{1}{3}$ it can tolerate $\gamma_{v} = 0.2$. There are only three parameters in the vertex expansion approach, as it does not require $\alpha$-solidarity, hence there is just one graph to show in this section.}
    \label{figure:vertex_interplay}
%
\end{figure}

Given a $d$-regular graph it can be shown that the inner boundary vertex expansion of the graph is at least as high as the graph conductance. Assume w.l.o.g that $|A|\leq|A^{c}|$, since $\partial_{v}(A,A^{c})\cdot d\geq e(A,A^{c})$ it follows that:
$$\frac{\partial_{v}(A,A^{c})}{|A|} = \frac{d\cdot\partial_{v}(A,A^{c})}{d\cdot|A|}\geq\frac{e(A,A^{c})}{vol(A)}$$

Going back to the numeric example in subsection \ref{subsection:edge_param_interplay}, now setting $\Phi_{v} = \frac{2}{5}$ then it follows that whenever $\frac{\gamma_{v}}{\beta}<\frac{2}{5}$, there exist a plethora of potential community histories for which a given community may potentially grow to be arbitrarily large. As an example, if the community wishes to achieve $\beta = 0.2$ then it can tolerate $\gamma_{v} = 0.08$. Figure~\ref{figure:vertex_interplay} illustrates the parameter interplay further. The line $\Phi_{v}=1$ shows a theoretical example where for each subset $A\subset V$, for every $x\in A$ there exist $y\in A^{c}$ such that $(x,y)\in E$. Assuming there is at least one honest identity in the community, and remembering that there cannot be an edge between an honest identity and a sybil identity, it follows that there are no sybils in any such community in $V$. The line $\Phi_{v}=1$ expresses this result as it shows that $\gamma_{v}=\beta$, which leads to $S=\emptyset$.

Maintaining $\Phi_{v} = 0.5$ leads to $\beta=2\gamma_{v}$ which means that the number of sybils in any such community is at most the number of corrupted identities that are willing to share an edge with a sybil identity. Unfortunately, the down side of using vertex expansion over conductance is that, as far as we know, there is no known way to measure or approximate vertex expansion better than the relation between vertex expansion and conductance shown above. We are also unaware of any method to construct a graph with vertex expansion $0.5$ or higher with a constant degree $d$.

\section{Outlook}\label{section:outlook}

We proposed two methods which allow a digital community to grow in a sybil-safe way. We analyzed them mathematically and showed that they are not only safe, but also feasible. Future research also includes mechanisms for penalizing the creation of attack edges while rewarding sybil hunting, modeling the possibility of honest identities abandoning the community, and using simulations to better understand the dynamics of safe growth.

\section*{Acknowledgements}

We thank the Braginsky Center for the Interface between Science and the Humanities for their generous support.

\bibliographystyle{plain}
\bibliography{bib}

\end{document}